# Spatially homogeneous models of Stäckel spacetimes of type (2.1)

*Evgeny Osetrin* *

*Konstantin Osetrin* **

*Altair Filippov* ***

*Tomsk State Pedagogical University*

## Abstract

All classes of spatially homogeneous space-time models are found that allow the integration of the equations of motion of test particles and the eikonal equation by the method of complete separation of variables according to type (2.1). Four classes of model data are obtained. The resulting models can be applied in any modified metric theories of gravity. Two of the above models allow solutions of the Einstein equations with a cosmological constant and radiation. For the models of a spatially homogeneous Universe with a cosmological constant and radiation obtained in Einstein's theory of gravity, the Hamilton-Jacobi equations of motion of the test particles and the eikonal equation for radiation are integrated by the method of separation of variables.

* Email: evgeny.osetrin@gmail.com
** Email: osetrin@tspu.edu.ru
*** Email: altair@tspu.edu.ru

# Spatially homogeneous models of Stäckel spacetimes of type (2.1)

*Evgeny Osetrin, Konstantin Osetrin, Altair Filippov*

Tomsk State Pedagogical University

## Introduction

As you know, spatially homogeneous models of space-time play a big role in building realistic models for the development of the Universe in any metric theory of gravitation. One of the important tools for studying such models is the study of geodetic lines in these spaces, including isotropic (light) ones. From this point of view, in the study of spatially homogeneous models of interest are the possibilities of analytical integration in these models of the eikonal equation and the equation of test particle motion in the Hamilton-Jacobi formalism by the method of variables separation.
Hamilton-Jacobi equation of the test particles in a gravitational field has the form:

$$g^{ij}\frac{\partial S}{\partial x^i}\frac{\partial S}{\partial x^j} = m^2, \qquad i,j = 0...3, \qquad (1)$$

Where $S = S(x^i)$ - action function of the test particle, $g^{ij}$ - metric of space-time, $m$ - mass of the test particle.

To avoid further confusion, we note that except for a capital letter $S$ we will also use a lowercase letter to indicate action function $s$ to denote the space-time interval.

The equation for radiation in a gravitational field has the form:

$$\tilde{g}^{ij}\frac{\partial \Psi}{\partial x^i}\frac{\partial \Psi}{\partial x^j} = 0, \qquad (2)$$

Where $\Psi = \Psi(x^i)$ - eikonal function.

Spaces that allow the existence of coordinate systems (CS) in which the Hamilton-Jacobi equations of the test particles in the form (1) are integrated by the method of complete separation of variables are called Stäckel (in honor of Paul Stäckel, see [1,2]).

Spaces admitting a complete separation of variables in the eikonal equation (2) are called conformal Stäckel spaces (SS).

According to the general theory of Stäckel spaces developed by V.N. Shapovalov [3,4], these spaces determined by the so-called "complete set" of Killing fields, consisting of the second rank Killing vectors and Killing tensors corresponding to a set of integrals of motion of test particles and meeting some algebraic conditions. A brief description of the main results of the Stäckel spaces theory can be found in [5].

The resulting spatially homogeneous models that allow integration of the eikonal equation (radiation) and the equation of test particles motion (dust matter) are of interest both for Einstein's classical theory of gravity and for other modified metric theories of gravity [6, 7, 8], including a comparative analysis of these models behavior in various modified theories of gravity.

In this work we will consider Stäckel spaces of type (2.1) admitting two Killing commuting vectors in the "complete set", therefore, in a privileged CS (where the separation of variables is allowed), the metric of the Stäckel space of type (2.1) can be written so that it depends only on two variables - $x^0$ and $x^1$ :

$$g^{ij} = \frac{1}{\Delta} \begin{pmatrix} 1 & 0 & 0 & 0 \\ 0 & 0 & f_1(x^1) & 1 \\ 0 & f_1(x^1) & A(x^0, x^1) & b_0(x^0) \\ 0 & 1 & b_0(x^0) & c_0(x^0) \end{pmatrix} \qquad (3)$$

Where $A(x^0, x^1) = a_0(x^0) + a_1(x^1)$, the function $\Delta$ is in the case of conformal Stäckel spaces an arbitrary function of all four variables, and in the case of Stäckel spaces of type (2.1) it has the form $\Delta = t_0(x^0) + t_1(x^1)$. Variables on which the metric in a privileged CS does not depend are called ignored (cyclic).

For Stäckel spaces, the action function $S$ of the test particles can be recorded in a privileged coordinate system in a "split" form. In the same coordinate system, it allows complete separation of variables and the eikonal equation, while the metric $\tilde{g}^{ij}$ will be different from $g^{ij}$ with the presence of an arbitrary conformal factor $\Delta$. For Stäckel spaces of type (2.1) (with metric (3) in the privileged CS) for the function of test particles action, we have (ignored variable enter linearly):

$$S = \phi_0(x^0) + \phi_1(x^1) + px^2 + qx^3, \qquad p, q, r - \text{const.} \qquad (4)$$

And, the functions $\phi_0(x^0)$ and $\phi_1(x^1)$ are the solutions of ordinary differential equations:

$$\dot{\phi}_0^2 = m^2 t_0(x^0) - p^2 a_0(x^0) - 2pq b_0(x^0) - q^2 c_0(x^0) - r, \qquad (5)$$

$$2\dot{\phi}_1 \left( pf_1(x^1) + q \right) = m^2 t_1(x^1) - p^2 a_1(x^1) + r, \qquad (6)$$

Where $m, p, q, r$ are the constant parameters of the test particles, and the dot means the ordinary derivative of the function of one variable.

Thus, in the considered space-time models, we can integrate with quadratures the equations of the test particles motion and radiation in the gravitational field. In the case of the eikonal equation, the vector $l_k = \partial \Psi / \partial x^k$ sets the radiation wave vector, and the equation $\Psi(x^i) = \text{const}$ sets the radiation wave surface (wave front).

**Classes of the spatially homogeneous Stäckel spaces (SS) models (2.1)**

When isolating spatially homogeneous models from the family of Stäckel spaces of type (2.1), we assume that the number of pairwise commuting Killing vectors of the considered models remain equal to two so that the metric in a privileged CS depends on two non-ignored variables, and commuting Killing vectors $X_0$ and $X_1$ from the complete set of SS (2.1) can be represented as:

$$X_0^i = (0, 0, 0, 1), \qquad X_1^i = (0, 0, 1, 0). \qquad (7)$$

Note that the vector $X_0$ is isotropic, and the vector $X_1$ - spatially similar.

Additional two Killing vectors providing spatial homogeneity of models have the form:

$$X_2 = \xi^i \partial_i, \qquad X_3 = \eta^j \partial_j \qquad (8)$$

The models under consideration, that allow the dependence of the metric on one of the non-ignored variables in the privileged CS only through the conformal factor, are assigned by us to the subtype B, in contrast to those where there is a dependence on both non-ignored variables are assigned by us to the subtype A. Models of type B refer to the intersection of the Stäckel space of type (2.1) and the set of conformal-Stäckel spaces of type (3.1) considered earlier in [9].

Spaces that allow the integration of the equations of the test particles motion by the method of the complete separation of variables in the Hamilton-Jacobi formalism make it possible to construct precisely integrable gravitational models both in Einstein's theory and in the modified theories of gravity with different types of matter [10] - [20].

For each A model type under consideration, the integration of the Einstein equations with the cosmological constant $\Lambda$ and the energy-momentum tensor of pure radiation with the energy density $\varepsilon$ and wave vector $l^k$ will be carried out below:

$$R_{ij} - \frac{1}{2} R g_{ij} = \Lambda g_{ij} + \varepsilon l_i l_j, \qquad l^k l_k = 0. \tag{9}$$

For models that correspond to Einstein's equations (9), the eikonal equation and the equation of test particles motion in the Hamilton-Jacobi form will also be integrated below.

**Model A.1 spatially homogeneous SS (2.1)**

For this class of the spatially homogeneous SS(2.1) we have:

$$f_1 = 0, \quad a_0 = 0, \quad A = a_1 = (x^1)^2, \quad b_0 = \alpha x^0, \quad c_0 = 0, \quad t_1 = 0, \quad \Delta = t_0 = 1/(x^0)^2, \quad \alpha \neq 0. \tag{10}$$

The metric and additional Killing vectors of the considered model can be written in the form:

$$g^{ij} = x^{0^2} \begin{pmatrix} 1 & 0 & 0 & 0 \\ 0 & 0 & 0 & 1 \\ 0 & 0 & (x^1)^2 & \alpha x^0 \\ 0 & 1 & \alpha x^0 & 0 \end{pmatrix}, \quad \xi^i = \begin{bmatrix} 0 \\ x^1 \\ x^2 \\ -x^3 \end{bmatrix}, \quad \eta^i = \begin{bmatrix} x^0 \\ 2x^1 \\ 3x^2 \\ 0 \end{bmatrix}, \tag{11}$$

Where $\alpha$ is a constant parameter of the model.

For the space-time interval, we have:

$$ds^2 = \frac{dx^{0^2}}{x^{0^2}} + \left(\alpha \frac{dx^1}{x^1} - \frac{dx^2}{x^0 x^1}\right)^2 + 2\frac{dx^1 dx^3}{x^{0^2}}. \tag{12}$$

Killing vector commutators of model A.1 have the form:
$[X_0, X_1] = 0, \quad [X_0, X_2] = -X_0, \quad [X_0, X_3] = 0, \quad [X_1, X_2] = X_1, \quad [X_1, X_3] = 3X_1, \quad [X_2, X_3] = 0,$

Where $X_0 = \delta_3^i, \quad X_1 = \delta_2^i, \quad X_2 = \xi^i, \quad X_3 = \eta^i$, moreover $X_1, X_2, X_3$ - the vector of the subgroup of spatial homogeneity. Sign-definiteness of the metric on the orbits of the spatial homogeneity subgroup imposes restrictions on the range of permissible values of the coordinates used: $x^1 x^3 < 0, x^{0^2} > 2|x^1 x^3|$.

Scalar curvature is constant and negative $R = -12$, Weyl tensor $C_{ijkl} \neq 0$. The model belongs to the type III according to the Bianchi classification and type N according to the Petrov classification.

If you introduce new variables $z = \ln x^0$ и $y = \ln x^1$, then the interval will take the form

$$ds^2 = dz^2 + \left(\alpha dy - e^{-(z+y)} dx^2\right)^2 + 2e^{(y-2z)} dy dx^3, \tag{13}$$

For metric (11), Einstein's equations (9) lead to the condition $\alpha = 0$, therefore, there are no field equations solutions for the model A.1.

**Model A.2 spatially homogeneous SS (2.1)**

For this class of the spatially homogeneous SS(2.1) we have:

$$f_1 = 0, \quad a_0 = 0, \quad A = a_1 = e^{2x^1}, \quad b_0 = 0, \quad c_0 = -\alpha^2(x^0)^2, \quad t_1 = 0, \quad \Delta = t_0 = 1/(x^0)^2. \tag{14}$$

$$g^{ij} = x^{0^2}\begin{pmatrix} 1 & 0 & 0 & 0 \\ 0 & 0 & 0 & 1 \\ 0 & 0 & e^{2x^1} & 0 \\ 0 & 1 & 0 & -\alpha^2 x^{0^2} \end{pmatrix}, \quad \xi^i = \begin{bmatrix} x^0 \\ -1 \\ 0 \\ 2x^3 \end{bmatrix}, \quad \eta^i = \begin{bmatrix} 0 \\ 1 \\ x^2 \\ 0 \end{bmatrix}, \tag{15}$$

Where $\alpha$ is a constant parameter of the model.

With $\alpha = \pm 1$ space admits three pairwise commuting Killing vectors and degenerates into a Stäckel space of type (3.1) - we do not consider this case. The metric property of having a fixed sign on the orbits of the spatial homogeneity group leads to restrictions on the range of allowed values of the coordinates used: $(1+\alpha^2)x^{0^2} - 4x^3 > 0, (\alpha^2 x^{0^4} - 4x^{3^2}) > 0.$ It follows that $\alpha \neq 0$.

The space-time interval has the form:

$$ds^2 = \frac{(dx^0)^2}{(x^0)^2} + \alpha^2(dx^1)^2 + 2\frac{dx^1 dx^3}{(x^0)^2} + \frac{e^{-2x^1}}{(x^0)^2}(dx^2)^2, \quad \alpha \neq 0, \pm 1. \tag{16}$$

Killing vector commutators of the model A.2 have the form:
$[X_0, X_1] = 0, \quad [X_0, X_2] = 2X_0, \quad [X_0, X_3] = 0, \quad [X_1, X_2] = 0, \quad [X_1, X_3] = X_1, \quad [X_2, X_3] = 0,$

Scalar curvature constant and negative $R = -12$, the components of the Weyl tensor are proportional $C_{ijkl} \sim (\alpha^2 - 1) \neq 0.$ The model belongs to the type III according to the Bianchi classification and the type N according to the Petrov classification.

The solution of Einstein's equations (9) for the metric (15) gives the following result:

$$\Lambda = 3, \quad l_0 = l_2 = l_3 = 0, \quad \varepsilon l_1^2 = \alpha^2 - 1, \quad |\alpha| > 1. \tag{17}$$

Thus we obtain a spatially homogeneous Universe with the cosmological constant $\Lambda$ filled with radiation with energy density $\varepsilon$ and wave vector $l^k = (0,0,0,l^3), l^3 = x^{0^2} l_1.$

**Integration of the Hamilton-Jacobi equation and the eikonal equation for the model A.2**

We integrate Hamilton-Jacobi equations (5) - (6) for the metric (15) and obtain the explicit form of the function $S$ for the test particle action (4) (when $q \neq 0$):

$$S(x^k) = m\ln(x^0) + \frac{1}{2}\left[\sqrt{m^2 + \alpha^2 q^2 x^{0^4} - rx^{0^2}} - m\ln\left(2m\sqrt{m^2 + \alpha^2 q^2 x^{0^4} - rx^{0^2}} + 2m^2 - rx^{0^2}\right)\right]$$
$$-\frac{r}{4\alpha q}\ln\left[2q\left(\sqrt{m^2 + \alpha^2 q^2 x^{0^4} - rx^{0^2}} + \alpha q x^{0^2}\right) - r\right] + \frac{1}{4q}\left(2rx^1 - p^2 e^{2x^1}\right) + px^2 + qx^3 + F(p,q,r), \tag{18}$$

Where $p, q, r$ - constant motion parameters of the test particles, $F(p,q,r)$ - an arbitrary function of constants.

When the test particle moves along the coordinate $x^0$ there may be turning points. When $r < 2|\alpha qm|$ there are no turning points, with $r = 2|\alpha qm|$ - there are two turning points with different signs and a "forbidden" zone along $x^0$ between them (infinite motion in $\pm\infty$ along $x^0$ from turning points of opposite signs) and with $r > 2|\alpha qm|$ - four turning points (two permitted driving zones along $x^0$ with different signs - in this case, we get two regions of finite motion of the test particle along $x^0$)

The eikonal function for the metric (15) (when $q \neq 0$) is in the following form:

$$\Psi = \frac{1}{2}\sqrt{\alpha^2 q^2 x^{0^4} - r x^{0^2}} - \frac{r}{4\alpha q}\ln\left[2\alpha q\left(\sqrt{\alpha^2 q^2 x^{0^4} - r x^{0^2}} + \alpha q x^{0^2}\right) - r\right]$$
$$+ \frac{1}{4q}\left(2rx^1 - p^2 e^{2x^1}\right) + px^2 + qx^3 + F(p,q,r).$$
(19)

When $r \leq 0$ there are no turning points along $x^0$. When $r > 0$ there are two turning points and a "forbidden" zone between them (infinite movement along $x^0$ in $\pm\infty$ from turning points of opposite signs). In the particular case when the motion constant $q$ vanishes, the equation (6) leads to conditions on constants

$$q = 0 \quad \rightarrow \quad p = r = 0.$$
(20)

Action when $q = p = r = 0$ takes the form

$$S = m \ln x^0.$$
(21)

If you enter a new variable $z = \ln x^0$, then for the interval we obtain ($x^3$ is an isotropic variable)

$$ds^2 = dz^2 + \alpha^2 dx^{1^2} + 2e^{-2z} dx^1 dx^3 + e^{-2(z+x^1)} dx^{2^2},$$
(22)

For the action function of the test particle in space-time with metric (22), we obtain (when $q \neq 0$)

$$S = mz + \frac{1}{2}\sqrt{m^2 - re^{2z} + q^2\alpha^2 e^{4z}} - \frac{m}{2}\ln\left[4m\left(\sqrt{m^2 - re^{2z} + q^2\alpha^2 e^{4z}} + m\right) - 2re^{2z}\right]$$
$$- \frac{r}{4q\alpha}\ln\left[2q\alpha\left(\sqrt{m^2 - re^{2z} + q^2\alpha^2 e^{4z}} + q\alpha e^{2z}\right) - r\right] + \frac{r}{2q}x^1 - \frac{p^2}{4q}e^{2x^1} + px^2 + qx^3 + F(p,q,r),$$
(23)

In the particular case when the constant of the test particle motion $q$ vanishes, for the particle action function we get $S = mz$ - the motion of a particle with a constant momentum along the coordinate $z$.

**Model A.3 spatially homogeneous SS (2.1)**

For this class of the spatially homogeneous SS(2.1) we have:
$$f_1 = 0, \quad a_0 = 0, \quad A = a_1 = \alpha \cos_T^{-1-\lambda}(x^1 - \beta)\sin_T^{-1+\lambda}(x^1 + \beta), \quad T = \pm 1,$$
$$b_0 = 0, \quad c_0 = Tx^{0^2}, \quad t_1 = 0, \quad \Delta = t_0 = 1/(x^0)^2.$$
(24)

$$g^{ij} = x^{0^2}\begin{pmatrix} 1 & 0 & 0 & 0 \\ 0 & 0 & 0 & 1 \\ 0 & 0 & \alpha\cos_T^{-1-\lambda}(x^1-\beta)\sin_T^{-1+\lambda}(x^1+\beta) & 0 \\ 0 & 1 & 0 & Tx^{0^2} \end{pmatrix},$$
(25)

$$\xi^i = \begin{bmatrix} x^0 \\ 0 \\ x^2 \\ 2x^3 \end{bmatrix}, \qquad \eta^i = \begin{bmatrix} x^0 \cos_{\tau} 2x^1 \\ \sin_{\tau} 2x^1 + \sin_{\tau} 2\beta \\ \gamma x^2 \\ \tau x^{0^2} \sin_{\tau} 2x^1 \end{bmatrix}, \qquad \lambda = \gamma / \cos_{\tau} 2\beta,$$

Permanent $\alpha, \beta$ и $\lambda$ - constant model parameters. When $\tau = -1$ harmonic functions are replaced by hyperbolic: $\sin_{(-1)}(x) = \mathrm{sh}(x), \cos_{(-1)}(x) = \mathrm{ch}(x)$.

If you enter a new variable $z = \ln x^0$, then for the interval we get:

$$ds^2 = dz^2 - \tau dx^{1^2} + e^{-2z}\left[ 2dx^1 dx^3 + (1/\alpha)\sin_{\tau}^{1-\lambda}(x^1 + \beta)\cos_{\tau}^{1+\lambda}(x^1 - \beta)\, dx^{2^2} \right], \tag{26}$$

Killing vector commutators of the model A.3 have the form:

$[X_0, X_1] = 0,\ [X_0, X_2] = 2X_0,\ [X_0, X_3] = 0,\ [X_1, X_2] = X_1,\ [X_1, X_3] = \gamma X_1,\ [X_2, X_3] = 0,$

Scalar curvature is constant and negative $R = -12$, all components of the Weyl tensor are proportional to the factor $(\lambda^2 - 1)\cos_{\tau}(2\beta)$. The model belongs to the type III according to the Bianchi classification and type N according to the Petrov classification.

Solving the Einstein equations (9) for the metric (25), we obtain:

$$\Lambda = 3, \qquad l_0 = l_2 = l_3 = 0, \qquad \tau l_1^2 = \frac{(1 - \lambda^2)\cos_{\tau}^2(2\beta)}{(\sin_{\tau}(2\beta) + \sin_{\tau}(2x^1))^2}, \qquad |\lambda| \le 1. \tag{27}$$

Thus we obtain a spatially homogeneous Universe with the interval (26) and with the cosmological constant $\Lambda$ filled with radiation with energy density $\varepsilon$ and wave vector $l^k = (0,0,0,l^3), l^3 = x^{0^2} l_1$.

When $\lambda = \pm 1$ and/or with $\cos(2\beta) = 0$ The Weyl tensor vanishes and the space-time model under consideration degenerates - it becomes a vacuum and conformally flat.

**Integration of the Hamilton-Jacobi equation and the eikonal equation for the model A.3 (harmonic functions in the metric)**

Integrating the Hamilton-Jacobi equation for $\tau = 1$ (harmonic functions $x^1$ in the metric), we obtain for the function of the action of the test particles (4) when $q \ne 0$:

$$S = \phi_0(x^0) + \phi_1(x^1) + px^2 + qx^3 + F(p, q, r), \qquad p, q, r - \mathrm{const}, \tag{28}$$

$$\phi_0(x^0) = m \ln x^0 + \frac{1}{2}\sqrt{m^2 - q^2 x^{0^4} - rx^{0^2}} - \frac{r}{4q}\arcsin\left(\frac{2q^2 x^{0^2} + r}{\sqrt{4m^2 q^2 + r^2}}\right)$$

$$-\frac{m}{2}\ln\left(2m^2 - rx^{0^2} + 2m\sqrt{m^2 - q^2 x^{0^4} - rx^{0^2}}\right) \tag{29}$$

$$0 < (x^0)^2 \le (x^0_{max})^2, \qquad (x^0_{max})^2 = \frac{r + \sqrt{r^2 + 4q^2 m^2}}{2q^2}. \tag{30}$$

Permanent $p, q, r$ are determined by the initial motion conditions of the test particle. We note that when the particle reaches the motion along the coordinate $x^0$ for the maximum value $x^0_{max}$ the rate of change of the test particles action function $x^0$ (zero component of the momentum ) changes sign. Thus $x^0_{max}$ - the turning point of the test particles motion along the coordinate $x^0$. $\phi_1(x^1)$ we get:

$$\phi_1(x^1) = \frac{r}{2q}x^1 - \frac{\alpha p^2}{2q\lambda \cos(2\beta)}\left(\frac{\sin(x^1+\beta)}{\cos(x^1-\beta)}\right)^\lambda \tag{31}$$

In the degenerate case when $q = 0$ from the equations of the test particle motion follows the condition $p = r = 0$ (since the function $a_1(x^1)$ is not constant). Then the function $\phi_0(x^0)$ takes a trivial form (movement of a particle with a constant momentum along $z$):

$$\phi_0 = m \ln x^0 = mz. \tag{32}$$

For the eikonal function, we obtain the expression ($r < 0$):

$$\Psi = \frac{1}{2}\sqrt{-rx^{0^2} - q^2 x^{0^4}} - \frac{r}{4q}\arcsin\left(\frac{2q^2 x^{0^2} + r}{\sqrt{4m^2 q^2 + r^2}}\right) + \frac{r}{2q}x^1 - \frac{\alpha p^2}{2q\lambda\cos(2\beta)}\left(\frac{\sin(x^1+\beta)}{\cos(x^1-\beta)}\right)^\lambda + F(p,q,r).$$

**Integration of the Hamilton-Jacobi equation and the eikonal equation for the model A.3 (hyperbolic functions in the metric)**

Integrating the Hamilton-Jacobi equation for $\tau = -1$ (hyperbolic functions $x^1$ in the metric), we get

$$\phi_0(x^0) = m \ln x^0 + \frac{1}{2}\sqrt{m^2 + q^2 x^{0^4} - rx^{0^2}} - \frac{r}{4q}\ln\left(m^2 + 2q^2 x^{0^2} + 2q\sqrt{m^2 + q^2 x^{0^4} - rx^{0^2}}\right)$$
$$- \frac{m}{2}\ln\left(2m^2 - rx^{0^2} + 2m\sqrt{m^2 + q^2 x^{0^4} - rx^{0^2}}\right), \tag{34}$$

When $r \geq 2|qm| > 0$ the motion of test particles in a variable $x^0$ has turning points $x^0_{min} \leq x^0 \leq x^0_{max}$.

$$\phi_1(x^1) = \frac{r}{2q}x^1 - \frac{\alpha p^2}{2q\lambda \text{ch}(2\beta)}\left(\frac{\text{sh}(x^1+\beta)}{\text{ch}(x^1-\beta)}\right)^\lambda \tag{35}$$

Then for the action function of the test particles we have (where $F(p,q,r)$ - the arbitrary function of parameters):

$$S = \phi_0(x^0) + \phi_1(x^1) + px^2 + qx^3 + F(p,q,r), \qquad p,q,r - \text{const}. \tag{36}$$

For the eikonal function, we obtain the expression:

$$\Psi = \frac{1}{2}\sqrt{q^2 x^{0^4} - rx^{0^2}} - \frac{r}{4q}\ln\left(qx^{0^2} + \sqrt{q^2 x^{0^4} - rx^{0^2}}\right) + \frac{rx^1}{2q} - \frac{\alpha p^2}{2q\lambda\text{ch}(2\beta)}\left(\frac{\text{sh}(x^1+\beta)}{\text{ch}(x^1-\beta)}\right)^\lambda + F(p,q,r).$$

(37)

**Model A.4 spatially homogeneous SS (2.1)**

This model is the only spatial homogeneous model when $f_1 \neq 0$. The metric and additional Killing vectors can be written as follows:

$$f_1 = x^1, \quad a_1 = 0, \quad A = a_0 = x^{0^2}, \quad b_0 = 0, \quad c_0 = 0, \quad t_1 = 0, \quad \Delta = t_0 = 1/(x^0)^2. \tag{38}$$

$$g^{ij} = x^{0^2} \begin{pmatrix} 1 & 0 & 0 & 0 \\ 0 & 0 & x^1 & 1 \\ 0 & x^1 & x^{0^2} & 0 \\ 0 & 1 & 0 & 0 \end{pmatrix}, \quad \xi^i = \begin{bmatrix} 0 \\ -x^1 \\ 0 \\ x^3 \end{bmatrix}, \quad \eta^i = \begin{bmatrix} x^0 \\ 2x^1 \\ 2x^2 \\ 0 \end{bmatrix}, \tag{39}$$

If you enter a variable $z = \ln x^0$, then the interval will take the form

$$ds^2 = dz^2 + 2e^{-2z} dx^1 dx^3 + e^{-4z} \left( dx^2 - x^1 dx^3 \right)^2, \tag{40}$$

Where $x^1$ is an isotropic wave variable.

Killing vector commutators of the model A.4 have the form:

$$[X_0, X_1] = 0, \ [X_0, X_2] = X_0, \ [X_0, X_3] = 0, \ [X_1, X_2] = 0, \ [X_1, X_3] = 2X_1, \ [X_2, X_3] = 0. \tag{41}$$

Sign-definiteness of the metric on the orbits of the spatial homogeneity subgroup impose restrictions on the range of permissible values of the coordinates used:
$x^1 x^3 < 0, \ x^{0^2} > 2|x^1 x^3|$.

Scalar curvature is constant and negative $R = -43/2$, Weyl tensor $C_{ijkl} \neq 0$. The model is of type D according to Petrov's classification and belongs to the type III according to Bianchi's classification.

There are no solutions of the Einstein equations with pure radiation (9) for the model A.4 with metric (39).

**Conclusion**

In this work, we classified spatially homogeneous space-time models that allow the existence of privileged coordinate systems for which the equation of test particles motion in the Hamilton-Jacobi form and the eikonal equation allow accurate integration by the method of complete separation of variables according to the type (2.1). The classification does not include subsets of spaces related to conformal-Stäckel spaces of type (3.1), which we obtained earlier in [6]. In total, 4 models of the type under consideration were obtained, which exhaust the classification.

All obtained models are of type III according to the Bianchi classification. Model A.4 refers to the type D according to Petrov's classification, the remaining models are of type N according to Petrov.
Model A.4 refers to wave-like models, i.e. to the space-time models which metrics in a privileged coordinate system (where the separation of variables is possible) depend on the isotropic wave variable.

The obtained models can be applied not only in Einstein's theory of gravity but also in other modified metric theories of gravity.

The obtained space-time models allow two exact solutions of the Einstein equations with the cosmological constant and the energy-momentum tensor of pure radiation, which are obtained explicitly (for two models, A.1 and A.4, there are no solutions to the Einstein equations with pure radiation).

For the obtained spatially homogeneous models of the Universe with radiation and the cosmological constant, an explicit form of the eikonal function and the form of complete integrals for the action function of test particles were found.

The study was carried out with the financial support of the Russian Federal Property Fund for research project No. 18-31-00040.

**List of references**